\documentclass[12pt]{article}

\usepackage{graphicx}
\begin{document}

\begin{center}
{\bf Magnetically charged black hole in framework of nonlinear electrodynamics model} \\
\vspace{5mm} S. I. Kruglov
\footnote{E-mail: serguei.krouglov@utoronto.ca}
\underline{}
\vspace{3mm}

\textit{Department of Chemical and Physical Sciences, University of Toronto,\\
3359 Mississauga Road North, Mississauga, Ontario L5L 1C6, Canada} \\
\vspace{5mm}
\end{center}
\begin{abstract}
A model of nonlinear electrodynamics is proposed and investigated in general relativity. We consider the magnetic black hole and find a regular solution which gives corrections into
the Reissner-Nordstr\"{o}m solution. At $r\rightarrow\infty$ the asymptotic spacetime becomes flat. The magnetic mass of the black hole is calculated and the metric function is obtained. At some values of the model parameter there can be one, two or no horizons. Thermodynamics of black holes is studied and we calculate the Hawking temperature and heat capacity of black holes. It is demonstrated that there is a phase transition of second order. At some parameters of the model black holes are thermodynamically stable.
\end{abstract}

\section{Introduction}

One of important problems of gravity physics is to develop a theory of quantum gravity. Till now we do not have a satisfactory and consistent theory of quantum gravity although the classical general relativity is the most successful theory of gravity. General relativity (GR) attract  much attention because the existence of black holes. In accordance with GR a high concentration of mass in space can result a strong gravitational field in such a way that even light cannot escape from that region named a black hole. There is an analogy between black hole physics and ordinary thermodynamics which could help to understand the theory of quantum gravity. According to Bardeen, Carter and Hawking \cite{bardeen} (see also \cite{Wald}) BH can be considered as a thermodynamic objects obeying four laws of black hole mechanics.
 It was understood that black holes can be considered as thermodynamic objects if quantum mechanical effects are combined with GR describing semiclassically BH physics. This allows us to study different thermodynamic properties of black holes: phase transitions, thermal stability, black hole evaporation and others.
The study of thermodynamic phase transitions of black holes can give quantum interpretation of GR and may help understanding the nature of quantum gravity. BH phase transitions were firstly investigated by Davies and Hut \cite{Davies}, \cite{Hut}.
Phase transitions play very important role in different branches of physics. Thermodynamic phase transitions of black holes can be studied by means of heat capacity in the canonical ensemble. When equilibrium is unstable the heat capacity becomes negative and the black hole absorbs a mass from outside. Then BH temperature decreases and the rate of absorption is greater than the rate of emission. As a result, the black hole will grow. The phase transitions take place when the discontinuities of heat capacity hold.
	
Models of nonlinear electrodynamics (NLED) coupled to GR are of interest because they
 can explain inflation of the universe \cite{Garcia}-\cite{Kruglov3}. The initial singularities in the early universe also can be avoided in some NLED models \cite{Novello1}. In particular NLED an upper limit on the electric field at the centre of point-like particles occurs \cite{Born}-\cite{Kruglov1}. In addition, in some models of NLED \cite{Born}-\cite{Kruglov2} the self-energy of charges is finite.
The self-energy of NLED is finite if a definite condition for the Lagrangian is satisfied \cite{Shabad2}, \cite{Shabad3}.
Due to loop corrections QED also becomes NLED \cite{Heisenberg}-\cite{Adler}.
Here we study a black hole within new model of NLED. Our model is a modification of the model given in \cite{Bronnikov}.
We investigate the model in the framework of GR and study thermodynamics of black holes.
In \cite{Bardeen}-\cite{Frolov} black holes were investigated in different NLED models.
At the weak field limit our NLED is transformed into Maxwell's electrodynamics and the correspondence principle takes place.

The structure of the paper is as follows.  We propose the new model of NLED with the parameter $\beta$ in section 2. In section 3 NLED in GR is studied. The asymptotics of the metric and mass functions at $r\rightarrow 0$ and $r\rightarrow\infty$ are found. We also obtained corrections to the Reissner-Nordstr\"{o}m (RN) solution. We have calculated the Hawking temperature and heat capacity of black holes. It was shown that  black holes undergo the second-order phase transition. The range where black holes are stable was obtained. In section 4 we made a conclusion.

The units with $c=\hbar=1$, $\varepsilon_0=\mu_0=1$ are used and the metric signature is given by $\eta=\mbox{diag}(-1,1,1,1)$.

\section{A model of NLED}

Let us study NLED with the Lagrangian density
\begin{equation}
{\cal L} = -\frac{{\cal F}}{\cosh\sqrt[4]{|\beta{\cal F}|}}.
 \label{1}
\end{equation}
The parameter $\beta$ possesses the dimensions of (length)$^4$, ${\cal F}=(1/4)F_{\mu\nu}F^{\mu\nu}=(\textbf{B}^2-\textbf{E}^2)/2$, and
$F_{\mu\nu}=\partial_\mu A_\nu-\partial_\nu A_\mu$ is the field tensor. The correspondence principle takes place because
at the weak field limit, $\beta {\cal F}\ll 1$, the Lagrangian density (1) is converted into Maxwell'l one, ${\cal L}\rightarrow-{\cal F}$.
From Eq. (1) one obtains field equations
\begin{equation}
\partial_\mu\left({\cal L}_{\cal F}F^{\mu\nu} \right)=0,
\label{2}
\end{equation}
where
\begin{equation}\label{3}
 {\cal L}_{\cal F}=\partial {\cal L}/\partial{\cal F}=\frac{|\beta{\cal F}|^{1/4}\sinh\sqrt[4]{|\beta{\cal F}|}}{4\cosh^2\sqrt[4]{|\beta{\cal F}|}}-\frac{1}{\cosh\sqrt[4]{|\beta{\cal F}|}}.
\end{equation}
Making use of Eq. (1) we find the electric displacement field $\textbf{D}=\partial{\cal L}/\partial \textbf{E}$
\begin{equation}
\textbf{D}=\varepsilon\textbf{E},~~~~\varepsilon=-{\cal L}_{\cal F},
\label{4}
\end{equation}
and the magnetic field $\textbf{H}=-\partial{\cal L}/\partial \textbf{B}$
\begin{equation}
\textbf{H}=\mu^{-1}\textbf{B},~~~~\mu^{-1}=-{\cal L}_{\cal F}=\varepsilon.
\label{5}
\end{equation}
With the help of Eqs. (4) and (5) the field equations (2) become the Maxwell equations
\begin{equation}
\nabla\cdot \textbf{D}= 0,~~~~ \frac{\partial\textbf{D}}{\partial
t}-\nabla\times\textbf{H}=0.
\label{6}
\end{equation}
The second pair of nonlinear Maxwell's equations, due to the Bianchi identity $\partial_\mu \tilde{F}^{\mu\nu}=0$, where $\tilde{F}^{\mu\nu}$ being the dual tensor, is
\begin{equation}
\nabla\cdot \textbf{B}= 0,~~~~ \frac{\partial\textbf{B}}{\partial
t}+\nabla\times\textbf{E}=0.
\label{7}
\end{equation}
 Eqs. (4) and (5) give the relationship
\begin{equation}
\textbf{D}\cdot\textbf{H}=(\varepsilon^2)\textbf{E}\cdot\textbf{B}.
\label{8}
\end{equation}
Thus, $\textbf{D}\cdot\textbf{H}\neq\textbf{E}\cdot\textbf{B}$ and according to \cite{Gibbons} the dual symmetry is broken.
The dual symmetry in classical electrodynamics and in Born-Infeld electrodynamics hold. It should be mentioned that in QED the dual symmetry is violated because of quantum corrections.

In NLED the symmetrical energy-momentum tensor is given by
\begin{equation}
T_{\mu\nu}=  {\cal L}_{\cal F}F_\mu^{~\alpha}F_{\nu\alpha}
-g_{\mu\nu}{\cal L}.
\label{9}
\end{equation}
By virtue of Eqs. (1), (3) and (9) we obtain the energy-momentum tensor trace
\begin{equation}\label{10}
 {\cal T}\equiv T^{\mu}_\mu=\frac{{\cal F}\sqrt[4]{|\beta{\cal F}|}\sinh\sqrt[4]{|\beta{\cal F}|}}{\cosh^2\sqrt[4]{|\beta{\cal F}|}},
\end{equation}
which is not zero and, as a result, the scale invariance is broken. At $\beta =0$ one comes to classical electrodynamics and the trace of the energy-momentum tensor, due to Eq. 10, vanishes.

\section{Magnetic black holes}

The action of NLED in general relativity is given by
\begin{equation}
I=\int d^4x\sqrt{-g}\left(\frac{1}{2\kappa^2}R+ {\cal L}\right),
\label{11}
\end{equation}
where $R$ is the Ricci scalar, $\kappa^2=8\pi G\equiv M_{Pl}^{-2}$, $G$ is Newton's constant,  and  $M_{Pl}$ is the reduced Planck mass.
We study magnetically charged black hole ($\textbf{E}=0$, $\textbf{B}\neq 0$).
Varying action (11) with respect to the metric and electric potential, one finds Einstein's and the electromagnetic field equations
\begin{equation}
R_{\mu\nu}-\frac{1}{2}g_{\mu\nu}R=-\kappa^2T_{\mu\nu},
\label{12}
\end{equation}
\begin{equation}
\partial_\mu\left(\sqrt{-g}{\cal L}_{\cal F}F^{\mu\nu}\right)=0.
\label{13}
\end{equation}
 The line element possessing the spherical symmetry is given by
\begin{equation}
ds^2=-f(r)dt^2+\frac{1}{f(r)}dr^2+r^2(d\vartheta^2+\sin^2\vartheta d\phi^2),
\label{14}
\end{equation}
where the metric function is defined as follows \cite{Bronnikov}:
\begin{equation}
f(r)=1-\frac{2GM(r)}{r}.
\label{15}
\end{equation}
The mass function is given by
\begin{equation}
M(r)=\int_0^r\rho_M(r)r^2dr,
\label{16}
\end{equation}
where $\rho_M$ is the magnetic energy density. The magnetic mass of the black hole is $m_M=\int_0^\infty\rho_M(r)r^2dr$.
From Eq. (9) we obtain the magnetic energy density (\textbf{E}=0)
\begin{equation}\label{17}
  \rho_M=T_0^{~0}=-{\cal L} = \frac{{\cal F}}{\cosh\sqrt[4]{|\beta{\cal F}|}},
\end{equation}
where ${\cal F}=B^2/2=q^2/(2r^4)$, and $q$ is a magnetic charge.

Introducing the dimensionless parameter $x=2^{1/4}r/(\beta^{1/4}\sqrt{q})$ and
making use of Eqs. (16) and (17) we obtain the mass function
\begin{equation}\label{18}
  M(x)=m_M-\frac{2^{1/4}q^{3/2}}{\beta^{1/4}}\arctan\left(\tanh\left(\frac{1}{2x}\right)\right),
\end{equation}
where the magnetic mass of the black hole reads
\begin{equation}\label{19}
 m_M = M(\infty)=\frac{\pi q^{3/2}}{2^{7/4}\beta^{1/4}}.
\end{equation}
With the aid of Eqs. (15) and (18) one obtains the metric function
\begin{equation}\label{20}
  f(x)=1-\frac{\pi-4\arctan\left(\tanh\left(\frac{1}{2x}\right)\right)}{ax},
\end{equation}
where $a=\sqrt{2\beta}/(Gq)$.
From Eq. (20) we find the asymptotic of the metric function at $r\rightarrow\infty$
\begin{equation}\label{21}
  f(r)=1-\frac{2Gm_M}{r}+\frac{Gq^2}{r^2}-\frac{G\sqrt{\beta}q^3}{6\sqrt{2} r^4}+
{\cal O}(r^{-6}).
\end{equation}
Equation (21) defines the corrections to the RN solution that are in the order of ${\cal O}(r^{-4})$.
When $r\rightarrow \infty$, $f(\infty)=1$, and the spacetime becomes Minkowski's spacetime.
We find that
\begin{equation}\label{22}
  \lim_{x\rightarrow 0^+}f(x)=1.
\end{equation}
In accordance with Eq. (22) the black hole is regular and it has not a conical singularity as $f(0)=1$.
At $\beta=0$ the model is converted into Maxwell's electrodynamics and Eq. (21) gives the RN solution.
The plot of the function $f(x)$ for different parameters $a=\sqrt{2\beta}/(Gq)$ is given in Fig. 1.
\begin{figure}[h]
\includegraphics[height=3.0in,width=3.0in]{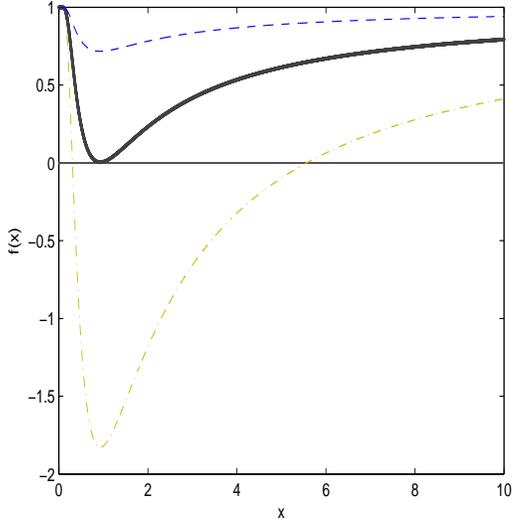}
\caption{\label{fig.1}The plot of the function $f(x)$. Dashed-dotted line corresponds to $a=0.5$, solid line corresponds to $a=1.42$ and dashed line corresponds to $a=5$.}
\end{figure}
Fig. 1 shows that at $a>1.42$ there are no horizons. If $a\simeq 1.42$ the extreme singularity occurs. At $a<1.43$ we have two horizons.
When $f(x_h)=0$ which defines horizons, we come from Eq. (20) to the equation
\begin{equation}\label{23}
 a=\frac{\pi-4\arctan\left(\tanh\left(\frac{1}{2x_h}\right)\right)}{x_h}.
\end{equation}
The plot of the function $a(x_h)$ is given in Fig. 2.
\begin{figure}[h]
\includegraphics[height=3.0in,width=3.0in]{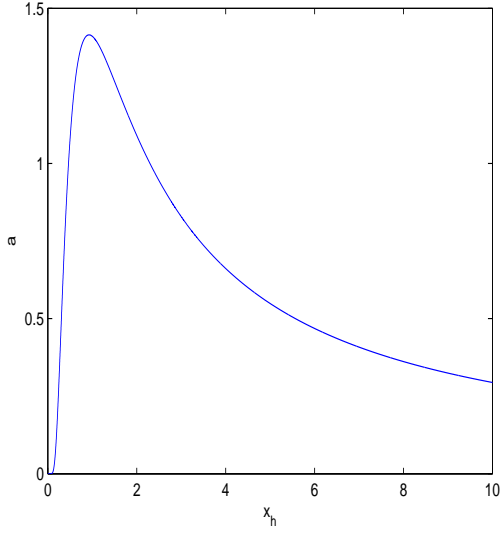}
\caption{\label{fig.2}The plot of the function $a(x_h)$.}
\end{figure}
The inner $x_-$ and outer $x_+$ horizons of the black hole are represented in Table 1.
\begin{table}[ht]
\caption{The inner and outer horizons of the black hole}
\centering
\begin{tabular}{c c c c c c c c c c }\\[1ex]
\hline \hline
$a$ & 0.4 & 0.5 & 0.6 & 0.7 & 0.8 & 0.9 & 1 & 1.3 & 1.4 \\[0.5ex]
\hline
$x_-$ & 0.280 & 0.307 & 0.334 & 0.363 & 0.394 & 0.428 & 0.466 & 0.648 & 0.816 \\[0.5ex]
\hline
$x_+$ & 7.158 & 5.569 & 4.503 & 3.731 & 3.145 & 2.680 & 2.297 & 1.402 & 1.065 \\[0.5ex]
\hline
\end{tabular}
\end{table}
At $r\rightarrow \infty$ the energy-momentum trace, due to Eqs. (10) (${\cal F}=q^2/(2r^4)$),
approaches zero. Therefore, according to Eq. (12) the Ricci scalar is $R=\kappa^2{\cal T}$, goes to zero, and spacetime becomes Minkowski's spacetime.

\subsection{Thermodynamics}

To study the thermal stability of magnetically charged black holes we will calculate the Hawking temperature. The expression for the Hawking temperature is given by
\begin{equation}
T_H=\frac{\kappa_S}{2\pi}=\frac{f'(r_h)}{4\pi}.
\label{24}
\end{equation}
Here $\kappa_S$ is the surface gravity and $r_h$ is the horizon. From Eqs. (15) and (16) one obtains the relations
\begin{equation}
f'(r)=\frac{2 GM(r)}{r^2}-\frac{2GM'(r)}{r},~~~M'(r)=r^2\rho_M,~~~M(r_h)=\frac{r_h}{2G}.
\label{25}
\end{equation}
Making use of Eqs. (17), (24) and (25), we obtain the Hawking temperature
\begin{equation}
T_H=\frac{1}{2^{7/4}\pi\beta^{1/4}\sqrt{q}} \left(\frac{1}{x_h}-\frac{2}{x_h^2\left[\pi-4\arctan(\tanh(1/2x_h))\right]\cosh(1/x_h)}\right).
\label{26}
\end{equation}
The plot of the function $T_H(x_h)$ is represented in Fig. 3.
\begin{figure}[h]
\includegraphics[height=3.0in,width=3.0in]{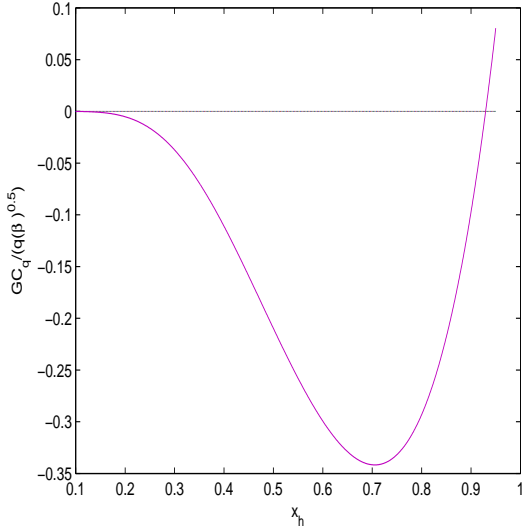}
\caption{\label{fig.3}The plot of the function $T_H\sqrt{q}\beta^{1/4}$ vs. horizons ($x_h$).}
\end{figure}
The Hawking temperature is zero at $x_h\simeq 0.93$. When $x_+>0.93$ the Hawking temperature is positive and for $x_-<0.93$ the Hawking temperature is negative. The Hawking temperature possesses the maximum at $x_+\simeq 1.82$ where
the heat capacity is singular and the second-order phase transition occurs in this point.
One can obtain the heat capacity from the relation
\begin{equation}
C_q=T_H\left(\frac{\partial S}{\partial T_H}\right)_q=\frac{T_H\partial S/\partial r_h}{\partial T_H/\partial r_h}=\frac{2\pi r_h T_H}{G\partial T_H/\partial r_h},
\label{27}
\end{equation}
where the entropy satisfies the Hawking area low $S=A/(4G)=\pi r_h^2/G$.
The plot of the function $GC_q/(\sqrt{\beta}q)$ vs. the horizon $x_h$ is given in Figs. 4 and 5 for different ranges of $x_h$.
\begin{figure}[h]
\includegraphics[height=3.0in,width=3.0in]{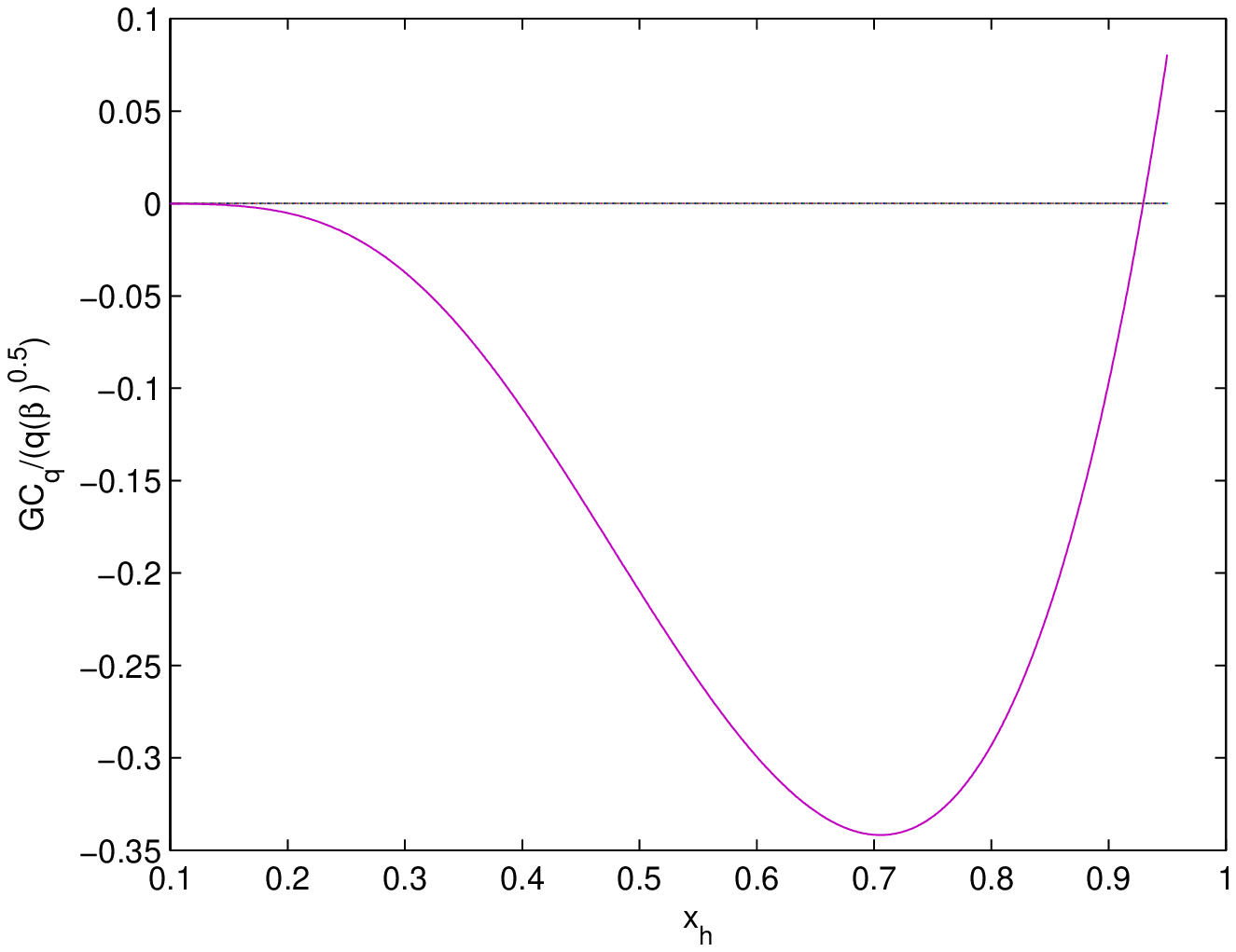}
\caption{\label{fig.4}The plot of the function $C_qG/(\sqrt{\beta}q)$ vs. $x_h$.}
\end{figure}
\begin{figure}[h]
\includegraphics[height=3.0in,width=3.0in]{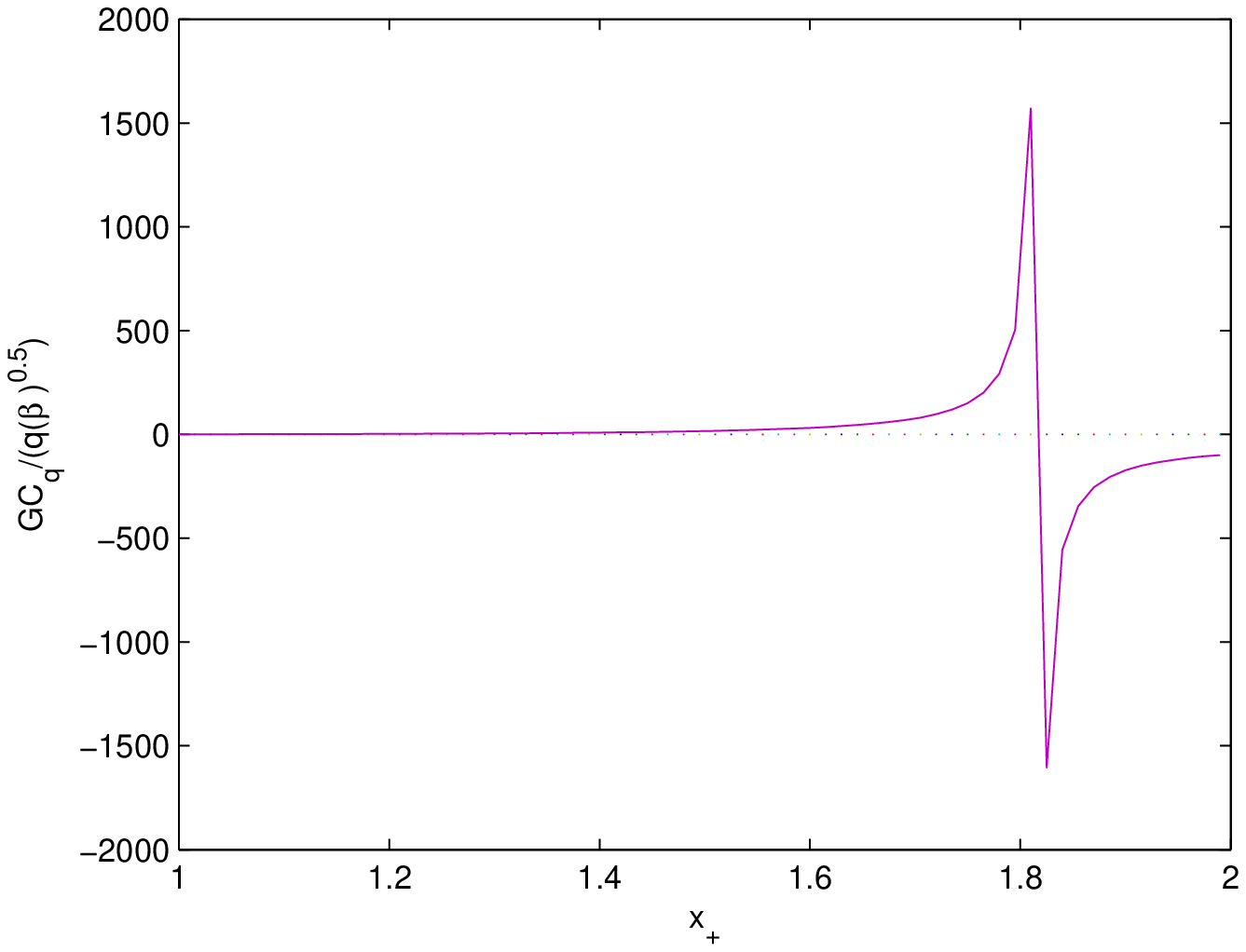}
\caption{\label{fig.5}The plot of the function $C_qG/(\sqrt{\beta}q)$ vs. $x_+$.}
\end{figure}
Figs. 4 and 5 show that indeed at $x_+\simeq 1.82$ ($r_+\simeq 1.53 \beta^{1/4}\sqrt{q}$), where heat capacity possesses discontinuity, the second-order phase transition of the black hole takes place. At $x_+<1.82$ the black hole is stable and at $x_+>1.82$ the heat capacity is negative and the black hole becomes unstable.

 \section{Conclusion}

We have proposed a NLED model with the dimensional parameter $\beta$ so that the model at weak field limit becomes Maxwell's electrodynamics. Thus, the correspondence principle takes place. NLED coupled with the gravitational field was investigated and regular black hole solution was obtained. We have studied the magnetized black holes and found the asymptotic of the metric and mass functions at $r\rightarrow 0$ and $r\rightarrow\infty$.
Corrections to the RN solution were obtained which are in the order of ${\cal O}(r^{-4})$. We have calculated the Hawking temperature and heat capacity of black holes and demonstrated that black holes undergo second-order phase transitions at $x_+\simeq 1.82$ ($r_+\simeq 1.53 \beta^{1/4}\sqrt{q}$). We have investigated the thermodynamic stabilities of black holes and shown that at the range $x_+<1.82$  the black holes are stable.
It is also interesting to study the perturbation stability of black holes in the framework of the model under consideration.

\end{document}